# Hardware-Efficient and Reliable Coherent DSCM Systems Enabled by Single-Pilot-Tone-Based Polarization Demultiplexing

Wei Wang, Dongdong Zou, Weihao Ni, and Fan Li

*Abstract*—Recently, coherent digital subcarrier multiplexing (DSCM) technology has become an attractive solution for next-generation ultra-high-speed datacenter interconnects (DCIs). To meet the requirements of low-cost and low-power consumption in DCI applications, a comprehensive simplification of the coherent DSCM system has been investigated. The pilot-tone-based polarization demultiplexing (PT-PDM) technique, known for its low-power consumption and ultra-fast polarization tracking capabilities, has emerged as a compelling alternative to the power-hungry *N*-tap adaptive multi-input multiple-output (MIMO) equalizer. However, the effectiveness of this PT-PDM technique is extremely vulnerable to the receiver-side XY-skew (Rx-XY-skew), which is revealed in this paper for the first time. Then, a pilot-tone-enabled modified Godard phase detector (PT-MGPD) scheme is proposed to realize Rx-XY-skew estimation, serving as the prerequisite for the successful implementation of the PT-PDM and simplification of the adaptive equalizer. Both the simulation and experiment are conducted to evaluate the accuracy of the proposed PT-MGPD scheme. The results prove it can achieve accurate estimation with an error of less than ±0.3ps. Besides, a low-complexity, high-spectral-efficiency, and ultra-fast polarization demultiplexing method based on a single pilot tone (SPT) is proposed for the DSCM system in this work. Based on the proposed PT-MGPD and SPT schemes, the conventional *N*-tap MIMO equalizer served for each subcarrier can be successfully pruned into two polarization-independent single-input single-output equalizers, and there is no performance penalty even if the polarization rotation speed reaches 10Mrad/s. According to the results, the proposed schemes provide a hardware-efficient and reliable coherent DSCM solution for next-generation ultra-high-speed DCIs.

## I. Introduction

WITH the development of new broadband services such as artificial intelligence, cloud computing, and ChatGPT, the data traffic raised from data centers has emerged as the predominate component of the total Internet traffic [1]. To meet the ever-growing demand for data transmission, a system with a large capacity is urgently desired for short-reach data center interconnects (DCIs). Although the intensity-modulation and direct-detection (IM/DD) system has always been regarded as the preferred solution for DCI applications, its application prospects in future ultra-high-speed transmission systems are expected to be quite challenging due to the limited available freedom for data delivery as well as its low reception linearity [2-4]. As an alternative solution, the coherent technology offering larger degrees of freedom for data transmission and better linearity has been widely discussed for next-generation DCI applications [5-8]. However, the conventional coherent system highly depends on the high-power consumption digital signal processing (DSP) algorithms, which is the major obstacle for coherent technology to substitute IM/DD systems in the cost- and power-sensitive DCI applications. To promote the application of coherent technology in short-reach scenarios, diminishing the computational complexity of the coherent DSP is essential [9-15].

Recently, the digital subcarrier multiplexing (DSCM) technique has received much attention in coherent systems, which provides the potential for the simplification of the chromatic dispersion compensation (CDC), alleviating the computational complexity of the receiver-side DSP (Rx-DSP) [16-18]. According to the evaluation conducted on the implementation of coherent detection for DCIs, the power consumption of the Rx-DSP is mainly determined by the adaptive equalizer [19]. Generally, a butterfly *N*-tap multiple-input multiple-output (MIMO) equalizer based on a constant modulus algorithm (CMA) or cascaded multi-modulus algorithm (CMMA) is indispensable in a dual-polarization coherent system to achieve polarization demultiplexing and equalization [20-23], while it results in high power consumption due to such butterfly MIMO operation. Particularly, in a DSCM system, each subcarrier is processed individually after subcarrier demultiplexing, implying that sophisticated MIMO operation is required for each subcarrier. To reduce the complexity of the equalizer, the conventional *N*-

Manuscript received XXX XXXX; revised XXX, XXXX; accepted XXX, XXXX. This work is partly supported by the National Key R&D Program of China (2023YFB2906000); National Natural Science Foundation of China (62271517, U2001601, 62035018); Guangdong Basic and Applied Basic Research Foundation (2023B1515020003), State Key Laboratory of Advanced Optical Communication Systems and Networks of China (2024GZKF19). (*Corresponding Author: Fan Li*)

W. Wang, D. Zou, Weihao Ni, and F. Li are with are with School of Electronics and Information Technology, the Guangdong Provincial Key Laboratory of Optoelectronic Information Processing Chips and Systems, Sun Yat-Sen University, Guangzhou 510275, China and Southern Marine Science and Engineering Guangdong Laboratory (Zhuhai), Zhuhai 519000, China (e-mail: lifan39@mail.sysu.edu.cn).

tap MIMO structure has been decomposed into two stages, i.e., a single-tap MIMO combined with two polarization-independent single-input single-output (SISO) equalizers [24, 25]. Since the effect of the polarization mode dispersion (PMD) can be ignored in short-reach scenarios, a single-tap 2×2 MIMO is sufficient to deal with the polarization crosstalk issue. In Ref. [26], an ultra-fast polarization demultiplexing scheme via three pilot tones is proposed, achieving simultaneous polarization de-crosstalk for all subcarriers before subcarrier demultiplexing. Such a pilot-tone-based polarization demultiplexing (PT-PDM) scheme not only leads to the complexity reduction of the subsequent equalizer but also improves the polarization tracking speed. Nevertheless, the algorithm proposed in Ref. [26] has an inherent performance limitation due to the arithmetic overflow when the amplitude of the received pilot tones on Y-polarization is close to zero. To address this issue, another joint polarization and carrier phase tracking algorithm based on two pilot tones is proposed in Ref. [27], which also reduces the required number of pilot tones. The scheme is then modified to be robust to the transmitter-side IQ impairments [28].

Such PT-PDM schemes indeed exhibit excellent performance in addressing polarization crosstalk with low complexity, further prompting the simplification of the subsequent equalizer for each subcarrier with a polarization-independent SISO structure. However, the effectiveness of these PT-PDM schemes will vanish in the presence of the receiver-side XY-skew (Rx-XY-skew) due to the incorrect Jones matrix calculation, which will be revealed and analyzed theoretically and experimentally in this paper for the first time. Therefore, calibrating the Rx-XY-skew is a prerequisite for implementing the PT-PDM scheme and simplifying the subsequent equalizer. Up to now, there has been scarce research on XY-skew estimation schemes in coherent systems. In Ref. [29], a transmitter-side XY-skew (Tx-XY-skew) alignment method based on reconfigurable interference is proposed, which is realized by sweeping different skew values for two tested tributaries carrying identical signals. Once the tested skew value is found, the amplitude of the resultant wave approaches zero. However, this method is only suitable for Tx-XY-skew detection and involves a rather complex operation. According to the analysis in Ref. [30], the Rx-XY-skew can be extracted from the tap coefficients of a 4×4 adaptive equalizer, while the estimation accuracy relies on the convergence of the equalizer and the complexity is quite high.

In this paper, a single-frequency-cosine pilot tone is used to achieve the Rx-XY-skew estimation for the DSCM system, which is enabled by a modified Godard phase detector (GPD) algorithm. Therefore, the proposed Rx-XY-skew estimation scheme is named the pilot-tone-enabled modified GPD (PT-MGPD). The verification of the proposed PT-MGPD scheme is conducted both in simulation and experiment. According to the results, the proposed PT-MGPD scheme shows robustness to the receiver-side IQ imbalance and can achieve the estimation error within ±0.3ps. Under the premise of implementing Rx-XY-skew calibration, these PT-PDM schemes can work effectively in decoupling polarization crosstalk, and the conventional adaptive MIMO equalizer served for each subcarrier can be subsequently pruned into two polarization-independent SISO equalizers, achieving a remarkable complexity reduction of the Rx-DSP. Furthermore, in order to eliminate the guard-band-induced spectral efficiency loss in the aforementioned multi-pilot-based polarization demultiplexing schemes, a modified low-complexity and high-spectral-efficiency polarization demultiplexing scheme enabled by a single pilot tone (SPT) is proposed in this paper. The effectiveness of the proposed SPT scheme is validated in dual-polarization 50Gbaud and 35Gbaud 4-subcarrier 16QAM (4SC-16QAM) DSCM systems by simulation and experiment respectively. The results indicate that the proposed SPT scheme can achieve ultra-fast polarization tracking after Rx-XY-skew calibration, and there is no performance penalty even with the polarization variation speed up to 10Mrad/s.

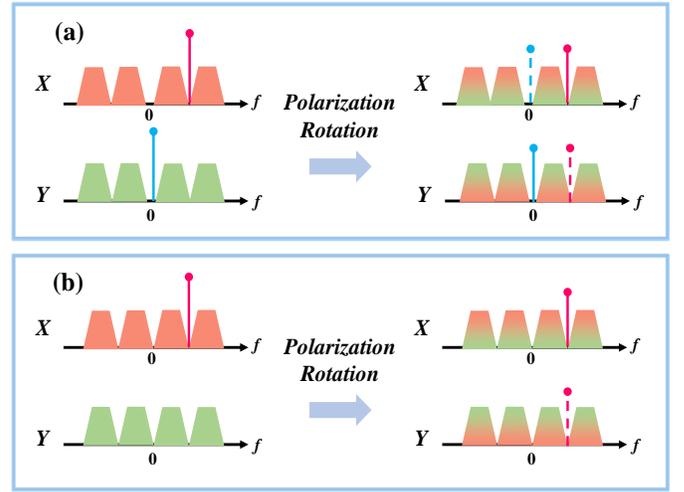

Fig. 1. Spectrums of DSCM signals with (a) two pilot tones and (b) a single pilot tone.

The rest of the paper is arranged as follows. In section II, the principles of the proposed SPT method as well as the dual pilot-tone (DPT) method proposed in Ref. [28] are introduced first. Then, the impact of the Rx-XY-skew on these PT-PDM schemes is analyzed theoretically. Finally, the algorithm principle of the proposed PT-MGPD method for Rx-XY-skew estimation is described in detail. In section III, the effectiveness of the proposed schemes is evaluated by simulation, and the performance penalty for the PT-PDM schemes induced by the Rx-XY-skew is also revealed. Furthermore, the experiment is conducted in section IV. Ultimately, a conclusion is drawn in section V.

## II. PRINCIPLE

### A. Operation principle of the DPT scheme

In the DPT scheme, two pilot tones with different frequencies are inserted at X-polarization and Y-polarization respectively as shown in Fig. 1(a). As described in Ref. [28], the transmitted signal together with the pilot tones can be expressed as:

$$\begin{bmatrix} E_X(t) \\ E_Y(t) \end{bmatrix} = \begin{bmatrix} S_X(t) + P_X(t) \\ S_Y(t) + P_Y(t) \end{bmatrix} \quad (1)$$

where $S_X(t)$ and $S_Y(t)$ represent the DSCM signals carried on X-polarization and Y-polarization, while $P_X(t)$ and $P_Y(t)$ denote the inserted pilot tones on two polarizations. Specifically, $P_X(t)$ and $P_Y(t)$ can be written as:

$$\begin{bmatrix} P_X(t) \\ P_Y(t) \end{bmatrix} = A \begin{bmatrix} \exp(j2\pi f_1 t) \\ \exp(j2\pi f_2 t) \end{bmatrix} \quad (2)$$

where $f_1$ and $f_2$ are the frequencies of two pilot tones respectively, and $A$ is the amplitude. Considering the polarization variation during fiber transmission, the received signal can be written as:

$$\begin{bmatrix} R_X(t) \\ R_Y(t) \end{bmatrix} = \begin{bmatrix} J_{XX}(t) & J_{XY}(t) \\ J_{YX}(t) & J_{YY}(t) \end{bmatrix} \begin{bmatrix} E_X(t) e^{j(2\pi\Delta f t+\varphi)} \\ E_Y(t) e^{j(2\pi\Delta f t+\varphi)} \end{bmatrix} + \begin{bmatrix} n_X(t) \\ n_Y(t) \end{bmatrix} \quad (3)$$

in which $J_{pp}(t)$ ($p \in [X,Y]$) is the parameter of the Jones matrix, $\Delta f$ and $\varphi$ are the frequency offset and phase noise, and $n_p(t)$ denotes the additive white Gaussian noise. By comparing the frequency of the received pilot tone to the transmitted one, the $\Delta f$ can be acquired and used for frequency offset compensation. Then, the down-converted signal can be expanded as:

$$\begin{bmatrix} R'_X(t) \\ R'_Y(t) \end{bmatrix} = \begin{bmatrix} J_{XX}(t)S_X(t) + J_{XY}(t)S_Y(t) \\ J_{YX}(t)S_X(t) + J_{YY}(t)S_Y(t) \end{bmatrix} e^{j\varphi} + A \begin{bmatrix} J_{XX}(t)e^{j2\pi f_1 t} + J_{XY}(t)e^{j2\pi f_2 t} \\ J_{YX}(t)e^{j2\pi f_1 t} + J_{YY}(t)e^{j2\pi f_2 t} \end{bmatrix} e^{j\varphi} + \begin{bmatrix} n_X(t) \\ n_Y(t) \end{bmatrix} \quad (4)$$

According to equation (4), the amplitudes of received pilot tones are related to the parameters of the Jones matrix. By down-converting these pilot tones to the baseband and filtering out, the extracted pilot tones can be expressed as:

$$\begin{bmatrix} P_{Xf_1}(t) \\ P_{Yf_1}(t) \end{bmatrix} = A \begin{bmatrix} J_{XX}(t) \\ J_{YX}(t) \end{bmatrix} e^{j\hat{\varphi}} + \begin{bmatrix} H\{n_X(t)\} \\ H\{n_Y(t)\} \end{bmatrix} \quad (5)$$

$$\begin{bmatrix} P_{Xf_2}(t) \\ P_{Yf_2}(t) \end{bmatrix} = A \begin{bmatrix} J_{XY}(t) \\ J_{YY}(t) \end{bmatrix} e^{j\hat{\varphi}} + \begin{bmatrix} H\{n_X(t)\} \\ H\{n_Y(t)\} \end{bmatrix} \quad (6)$$

where $H\{\cdot\}$ refers to a low pass filter operation. The Jones matrix reflecting the variation of the polarization state is contained in the filtered pilot tones with phase noise $\hat{\varphi}$. Therefore, the estimated Jones matrix is:

$$\begin{bmatrix} \hat{J}_{XX}(t) & \hat{J}_{XY}(t) \\ \hat{J}_{YX}(t) & \hat{J}_{YY}(t) \end{bmatrix} = \begin{bmatrix} J_{XX}(t)e^{j\hat{\varphi}} & J_{XY}(t)e^{j\hat{\varphi}} \\ J_{YX}(t)e^{j\hat{\varphi}} & J_{YY}(t)e^{j\hat{\varphi}} \end{bmatrix} \quad (7)$$

Once the Jones matrix is obtained, the polarization demultiplexing can be achieved by implementing the inverse Jones matrix as:

$$\begin{bmatrix} \hat{E}_X(t) \\ \hat{E}_Y(t) \end{bmatrix} = \begin{bmatrix} \hat{J}_{XX}(t) & \hat{J}_{XY}(t) \\ \hat{J}_{YX}(t) & \hat{J}_{YY}(t) \end{bmatrix}^{-1} \begin{bmatrix} R_X(t) \\ R_Y(t) \end{bmatrix} \quad (8)$$

$$\begin{bmatrix} \hat{J}_{XX}(t) & \hat{J}_{XY}(t) \\ \hat{J}_{YX}(t) & \hat{J}_{YY}(t) \end{bmatrix}^{-1} = \begin{bmatrix} J_{XX}(t) & J_{XY}(t) \\ J_{YX}(t) & J_{YY}(t) \end{bmatrix}^{-1} e^{-j\hat{\varphi}} \quad (9)$$

### B. Operation principle of the SPT scheme

It is worth noting that the Jones matrix is a unitary matrix as its elements have the following relationship [26, 31]:

$$\begin{aligned} J_{XX}(t) &= J^*_{YY}(t) \\ J_{XY}(t) &= -J^*_{YX}(t) \end{aligned} \quad (10)$$

where * refers to the conjugate operation. Taking advantage of this relationship, the Jones matrix can be acquired by only utilizing equation (5), meaning that a single pilot tone is sufficient to achieve polarization tracking. Fig. 1(b) gives the spectrum diagram of the SPT scheme. Only a single pilot tone is inserted in the X-polarization, and it will produce a project in the Y-polarization after polarization rotation as shown in Fig. 1(b). By extracting the $P_{Xf_1}$ and $P_{Yf_1}$ in two polarizations as equation (5), the Jones matrix can be obtained consequently as:

$$\begin{bmatrix} \hat{J}_{XX}(t) & \hat{J}_{XY}(t) \\ \hat{J}_{YX}(t) & \hat{J}_{YY}(t) \end{bmatrix} = \begin{bmatrix} J_{XX}(t)e^{j\hat{\varphi}} & -J^*_{YX}(t)e^{-j\hat{\varphi}} \\ J_{YX}(t)e^{j\hat{\varphi}} & J^*_{XX}(t)e^{-j\hat{\varphi}} \end{bmatrix}$$
$$= \begin{bmatrix} J_{XX}(t)e^{j\hat{\varphi}} & J_{XY}(t)e^{-j\hat{\varphi}} \\ J_{YX}(t)e^{j\hat{\varphi}} & J_{YY}(t)e^{-j\hat{\varphi}} \end{bmatrix} \quad (11)$$

Then, the polarization demultiplexing can be carried out by:

$$\begin{bmatrix} \hat{E}_X(t) \\ \hat{E}_Y(t) \end{bmatrix} = \begin{bmatrix} J_{XX}(t)e^{j\hat{\varphi}} & J_{XY}(t)e^{-j\hat{\varphi}} \\ J_{YX}(t)e^{j\hat{\varphi}} & J_{YY}(t)e^{-j\hat{\varphi}} \end{bmatrix}^{-1} \begin{bmatrix} R_X(t) \\ R_Y(t) \end{bmatrix}$$
$$= \begin{bmatrix} e^{-j\hat{\varphi}} & 0 \\ 0 & e^{j\hat{\varphi}} \end{bmatrix} \begin{bmatrix} J_{XX}(t) & J_{XY}(t) \\ J_{YX}(t) & J_{YY}(t) \end{bmatrix}^{-1} \begin{bmatrix} R_X(t) \\ R_Y(t) \end{bmatrix} \quad (12)$$

It is worth noting that the influence of the phase noise $\hat{\varphi}$ contained in the estimated Jones matrix can be addressed by the blind phase search (BPS) algorithm during carrier phase recovery for both the DPT and SPT schemes.

### C. Impact of the Rx-XY-skew

As described above, the polarization crosstalk can be eliminated by implementing the calculated inverse Jones matrix based on the DPT or SPT scheme. However, in a practical receiver, the path lengths of four tributaries may not perfectly match, which introduces not only the receiver-side IQ skew but also the Rx-XY-skew. Unfortunately, the Rx-XY-skew will seriously deteriorate the effectiveness of these PT-PDM schemes. In the presence of the Rx-XY-skew, the filtered pilot

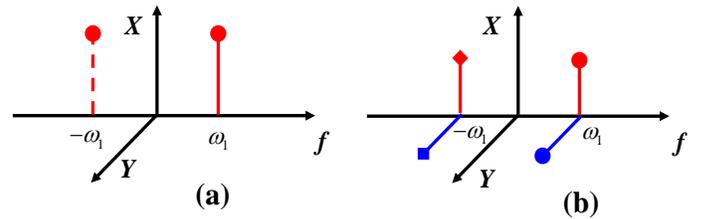

Fig. 2. (a) Training signal of the proposed PT-MGPD scheme. (b) Pilot tones after polarization rotation.

tones can be represented as:

$$\begin{bmatrix} P_{Xf_1}(t) \\ P_{Yf_1}(t) \end{bmatrix} = A \begin{bmatrix} J_{XX}(t) \\ J_{YX}(t-\tau_{XY})e^{-j2\pi f_1 \tau_{XY}} \end{bmatrix} e^{j\hat{\varphi}} + \begin{bmatrix} H\{n_X(t)\} \\ H\{n_Y(t)\} \end{bmatrix} \quad (13)$$

$$\begin{bmatrix} P_{Xf_2}(t) \\ P_{Yf_2}(t) \end{bmatrix} = A \begin{bmatrix} J_{XY}(t) \\ J_{YY}(t-\tau_{XY})e^{-j2\pi f_2 \tau_{XY}} \end{bmatrix} e^{j\hat{\varphi}} + \begin{bmatrix} H\{n_X(t)\} \\ H\{n_Y(t)\} \end{bmatrix} \quad (14)$$

where $\tau_{XY}$ is the Rx-XY-skew. In this context, the estimated Jones matrix based on the DPT scheme is expressed as:

$$JM_{DPT} = \begin{bmatrix} J_{XX}(t) & J_{XY}(t) \\ J_{YX}(t-\tau_{XY})e^{-j2\pi f_1 \tau_{XY}} & J_{YY}(t-\tau_{XY})e^{-j2\pi f_2 \tau_{XY}} \end{bmatrix} \quad (15)$$

while the estimated Jones matrix based on the SPT scheme can be written as:

$$JM_{SPT} = \begin{bmatrix} J_{XX}(t) & -J_{YX}^*(t-\tau_{XY})e^{j2\pi f_1 \tau_{XY}} \\ J_{YX}(t-\tau_{XY})e^{-j2\pi f_1 \tau_{XY}} & J_{XX}^*(t) \end{bmatrix} \quad (16)$$

Obviously, whether in the DPT or SPT scheme, the Rx-XY-skew will lead to an incorrect estimation of the Jones matrix. Therefore, it will seriously deteriorate the performance of polarization demultiplexing. Nevertheless, it is not an issue in the conventional CMMA equalization method, since the impact of the Rx-XY-skew can be addressed by the *N*-tap MIMO operation.

### D. Principle of the proposed PT-MGPD method

According to the analysis in section II-C, it is crucial to address the Rx-XY-skew issue before implementing the hardware-efficient PT-PDM scheme along with the simplification of the subsequent equalizer. In this work, a low-complexity PT-MGPD method is proposed to achieve the Rx-XY-skew estimation, which is enabled by a single-frequency-cosine pilot tone only located in X-polarization. Fig. 2(a) gives the structure of the training signal in the frequency domain. A cosine pilot tone with frequency $f_1$ is inserted in the X-polarization. The pilot tone represented by the dashed line in negative frequency refers to the conjugate component of the positive-frequency one. Therefore, the inserted pilot tone signal can be written as:

$$PX_{f_1}(t) = A\cos(2\pi f_1 t) \quad (17)$$

where A is the amplitude of the pilot tone signal. Since the designed training pilot-tone signal is a real signal with X-polarization, it only suffers the transmitter-side time delay of the I-tributary in X-polarization, thus can be represented as:

$$PX'_{f_1}(t) = A\cos(2\pi f_1 (t-\tau_{TXI})) \quad (18)$$

where $\tau_{TXI}$ is the time delay of the I-tributary in X-polarization at the transmitter. It is worth noting that the proposed estimation method is designed to be done in an optical back-to-back (OBTB) self-coherent configuration. Therefore, only the polarization rotation is considered, while the effects of CD, PMD, and frequency offset are ignored here. The pilot tone signal with polarization rotation can be expressed as:

$$\begin{bmatrix} ae^{j\alpha} & be^{j\beta} \\ -be^{-j\beta} & ae^{-j\alpha} \end{bmatrix} \begin{bmatrix} PX'_{f_1}(t) \\ 0 \end{bmatrix} = \begin{bmatrix} ae^{j\alpha}\cos(2\pi f_1(t-\tau_{TXI})) \\ -be^{-j\beta}\cos(2\pi f_1(t-\tau_{TXI})) \end{bmatrix}$$

(19)

where $a$ ($b$) and $\alpha$ ($\beta$) are the amplitude and phase of $J_{XX}$ ($J_{XY}$), and the amplitude $A$ in equation (18) is normalized to 1. Fig. 2(b) shows the pilot tones after polarization rotation. It can be observed that the pilot tones will project onto the other polarization due to polarization rotation, and the pilot tone in the negative frequency is no longer the conjugate component of the positive-frequency one. Then, the pilot tones in X-polarization and Y-polarization can be denoted as:

$$PX''_{f_1}(t) = ae^{j\alpha}\cos(2\pi f_1(t-\tau_{TXI})) \quad (20)$$

$$PY''_{f_2}(t) = -be^{-j\beta}\cos(2\pi f_1(t-\tau_{TXI})) \quad (21)$$

When considering the path length mismatch at the receiver, each tributary will experience different time delays, which can be represented as:

$$PXI''_{f_1}(t) = a\cos\alpha\cos(2\pi f_1(t-\tau_{TXI}-\tau_{RXI})) \quad (22)$$

$$PXQ''_{f_1}(t) = a\sin\alpha\cos(2\pi f_1(t-\tau_{TXI}-\tau_{RXQ})) \quad (23)$$

$$PYI''_{f_1}(t) = -b\cos\beta\cos(2\pi f_1(t-\tau_{TXI}-\tau_{RYI})) \quad (24)$$

$$PYQ''_{f_1}(t) = -b\sin\beta\cos(2\pi f_1(t-\tau_{TXI}-\tau_{RYQ})) \quad (25)$$

where $\tau_{RXI}$, $\tau_{RXQ}$, $\tau_{RYI}$, and $\tau_{RYQ}$ denote the time delays in XI, XQ, YI, and YQ tributaries at the receiver. Generally, the Rx-XY-skew is defined as:

$$\tau_{RXY} = \tau_{RYI} - \tau_{RXI} \quad (26)$$

It can be observed that equation (23) and equation (25) are related to $\tau_{RXI}$ and $\tau_{RYI}$ respectively. Then, transforming equation (23) and equation (25) into the frequency domain, we

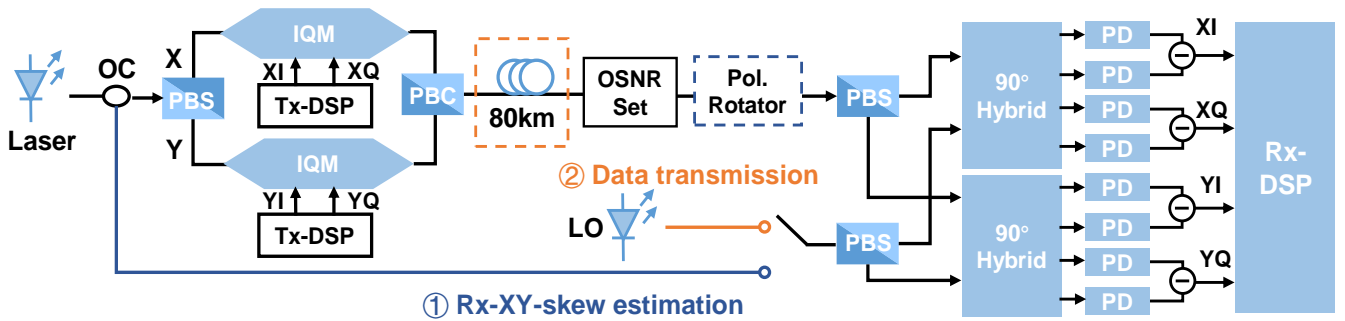

Fig. 3. Optical transmission system established in VPI for simulation.

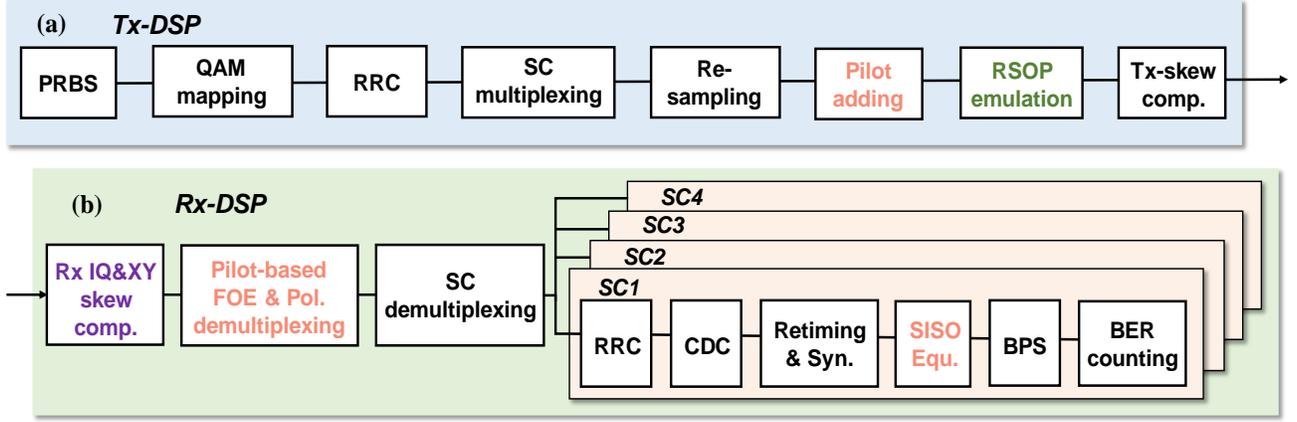

Fig. 4. (a) Transmitter-side and (b) receiver-side DSP flows.

can obtain:

$$PXI_{f_1}^{"}(f) = \frac{a\cos\alpha}{4}\left[e^{-j2\pi f_1(\tau_{TXI}+\tau_{RXI})}\delta(f+f_1) + e^{j2\pi f_1(\tau_{TXI}+\tau_{RXI})}\delta(f-f_1)\right] \quad (27)$$

$$PYI_{f_1}^{"}(f) = \frac{-b\cos\beta}{4}\left[e^{-j2\pi f_1(\tau_{TXI}+\tau_{RYI})}\delta(f+f_1) + e^{j2\pi f_1(\tau_{TXI}+\tau_{RYI})}\delta(f-f_1)\right] \quad (28)$$

where $\delta(\cdot)$ is the impulse response function. According to equation (28) and equation (29), the time delays result in phase shifts in the frequency domain, which are contained in the phase information. By implementing the conjugate multiplication and angle extraction of $PXI_{f_1}^{"}(f)$ and $PYI_{f_1}^{"}(f)$ as:

$$angle\left(PXI_{+f_1}^{"} \cdot PXI_{-f_1}^{"*}\right) = -4\pi f_1(\tau_{TXI}+\tau_{RXI}) \quad (29)$$

$$angle\left(PYI_{+f_1}^{"} \cdot PYI_{-f_1}^{"*}\right) = -4\pi f_1(\tau_{TXI}+\tau_{RYI}) \quad (30)$$

where $PXI_{+f_1}^{"}$ ($PYI_{+f_1}^{"}$) and $PXI_{-f_1}^{"}$ ($PYI_{-f_1}^{"}$) represent the positive and negative pilot tone components of $PXI_{f_1}^{"}$ ($PYI_{f_2}^{"}$). According to equation (30) and equation (31), the Rx-XY-skew $\tau_{XY}$ can be calculated as:

$$\tau_{XY} = \frac{1}{4\pi f_1}\left[angle\left(PXI_{+f_1}^{"} \cdot PXI_{-f_1}^{"*}\right) - angle\left(PYI_{+f_1}^{"} \cdot PYI_{-f_1}^{"*}\right)\right] \quad (31)$$

Once the Rx-XY-skew is estimated, the estimated value is used for Rx-XY-skew calibration during data transmission. It is worth noting that a certain degree of polarization crosstalk is necessary to achieve the estimation of the Rx-XY-skew in the proposed scheme. Although only the Rx-XY-skew is estimated in this paper, the proposed skew estimation scheme can also be further extended to a version that can achieve the transceiver IQ-skew and XY-skew estimation by appropriately adding certain pilot tones.

## III. SIMULATION DISCUSSION

In this section, the simulation is conducted to investigate the performance of the DPT and SPT methods, the impact of the Rx-XY-skew, and the effectiveness of the proposed PT-MGPD method.

### A. Simulation setup

The simulation is carried out by MATLAB and VPItransmissionmaker (VPI) software. Specifically, the transceiver DSP algorithms are realized by MATLAB, while the electro-optical/optical-electro (E-O/O-E) conversion and optical transmission system are established in VPI as shown in Fig. 3. At the transmitter side, the optical carrier emitting from a laser with linewidth and output power of 100kHz and 16dBm is divided into two orthogonal polarizations by a polarization beam splitter (PBS) and then injected into two IQ modulators (IQMs). Signals output from two IQMs are coupled by a polarization beam combiner (PBC) and launched into an 80km fiber. An optical signal-to-noise (OSNR) setting module is implemented to adjust the OSNR of the system. At the receiver side, both the incoming signal and the local oscillator (LO) are divided into X-polarization and Y-polarization, which are injected into two 90° optical hybrids. After that, the signals are detected by four pairs of photodiodes (PDs), and the detected signals are processed in the Rx-DSP. Due to the lack of the polarization scrambler module to generate variable polarization rotation speed, the rotation of the state of polarization (RSOP) is emulated by DSP. As for the Rx-XY-skew estimation, a self-coherent configuration operating in OBTB is required. In this case, the transmitter-side laser is used as the LO. Besides, a polarization rotation module is added in the signal transmission link to introduce the polarization crosstalk, which is essential for Rx-XY-skew estimation.

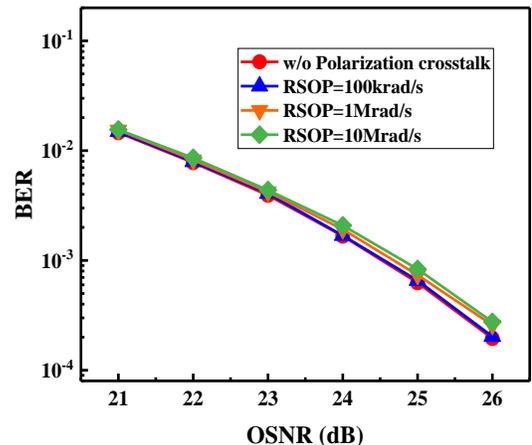

Fig. 5. BER versus OSNR of the SPT method under different RSOP speeds.

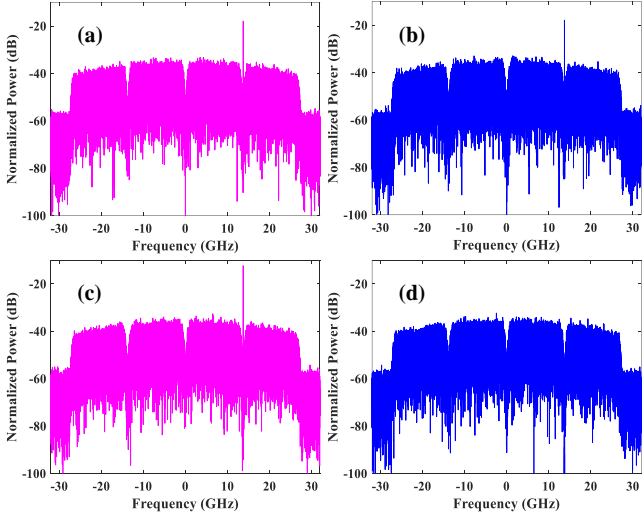

Fig. 6. Spectrums of the received DSCM signals before polarization demultiplexing (a) X-polarization, (b) Y-polarization. Spectrums of the received DSCM signals after polarization demultiplexing (c) X-polarization, (d) Y-polarization.

The transmitter-side DSP (Tx-DSP) and Rx-DSP for data transmission are given in Figs. 4(a) and 4(b) respectively. In the Tx-DSP, the pseudo-random binary sequence (PRBS) is mapped into 16QAM symbols. The generated QAM symbols are shaped by a root-raised cosine (RRC) filter with a roll-off factor of 0.1. Then, the shaped signals are multiplexed to generate the 4SC-16QAM DSCM signals for two polarizations. After resampling, a pilot tone is added to the X-polarization, which is simultaneously used for frequency offset estimation (FOE) and polarization demultiplexing. Subsequently, the RSOP is emulated digitally. In a practical system, the transmitter-side IQ skew (Tx-skew) should be calibrated before digital-to-analog conversion, while it can be ignored in the simulation. As for the Rx-DSP, after receiver-side IQ skew (Rx-skew) and Rx-XY-skew compensation, the pilot-tone-based FOE and polarization demultiplexing are performed. Then, the subcarrier demultiplexing, RRC filter, CDC, retiming, and synchronization are conducted in order. Since the polarization crosstalk is eliminated before the subcarrier demultiplexing, two polarization-independent SISO equalizers are implemented for each subcarrier followed by the blind phase search (BPS) algorithm. Finally, the bit error ratio (BER) is calculated.

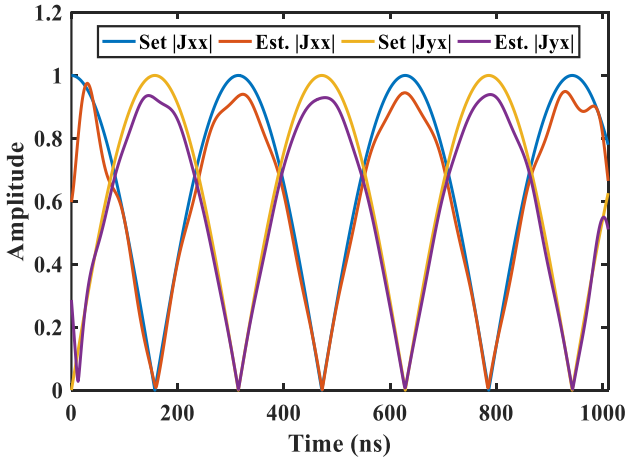

Fig. 7. Amplitude of the estimated Jones matrix.

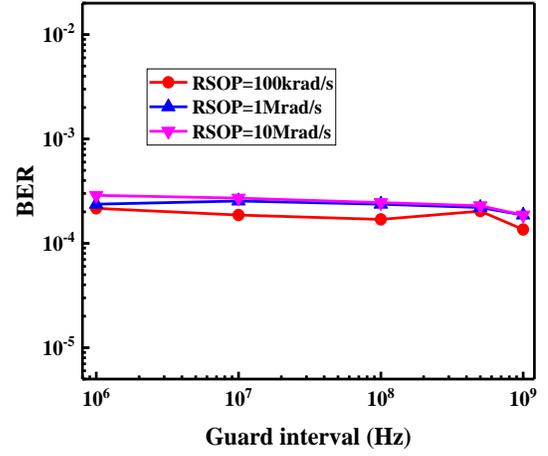

Fig. 8. BER performance of the SPT method with different frequency intervals between two subcarriers.

To emulate the RSOP in the transmission system, the polarization model is established as [32]:

$$JM(f) = \begin{bmatrix} \sqrt{1+\gamma} & 0 \\ 0 & \sqrt{1-\gamma} \end{bmatrix} \times \begin{bmatrix} \cos\alpha e^{j\beta} & -\sin\alpha e^{j\eta} \\ \sin\alpha e^{-j\eta} & \cos\alpha e^{-j\beta} \end{bmatrix} \\ \times \begin{bmatrix} e^{j\pi f\tau} & 0 \\ 0 & e^{-j\pi f\tau} \end{bmatrix} \quad (32)$$

The first matrix in equation (33) refers to the polarization-dependent loss (PDL) matrix, where $\gamma$ is the magnitude of the PDL. It can be defined in dB as:

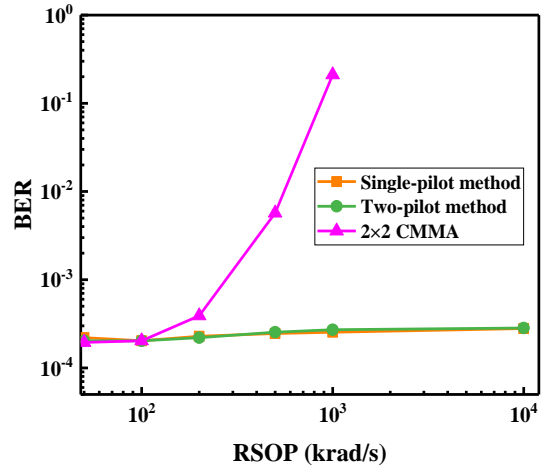

Fig. 9. BER versus RSOP of SPT, DPT, and the conventional N-tap 2×2 CMMA methods.

$$\Upsilon(dB) = 10\log_{10}\frac{1+\gamma}{1-\gamma} \quad (33)$$

The second matrix represents the RSOP, where $\alpha$, $\beta$, and $\eta$ are the parameters describing the polarization rotation. Supposing that the polarization state is varying at a certain speed, the values of $\alpha$, $\beta$, and $\eta$ can be considered as changing with an increment $\Delta(k)$:

$$\alpha(k) = \alpha_0 + \Delta(k) \quad (34)$$
$$\beta(k) = \beta_0 + \Delta(k) \quad (35)$$
$$\eta(k) = \eta_0 + \Delta(k) \quad (36)$$

where $\alpha_0$, $\beta_0$, and $\eta_0$ are the initial phase, and $k$ is the sample index. The increment $\Delta(k)$ can be defined as:

$$\Delta(k) = \frac{\Omega k}{f_s} \quad (37)$$

where $\Omega$ and $f_s$ are the RSOP speed and sampling rate, respectively.

The third matrix is the PMD matrix, in which the $\tau$ refers to the differential group delay (DGD). However, the effect of the PMD can be ignored in the short-reach scenarios, and it is not considered in this paper.

*B. Performance analysis of the SPT scheme*

First, the effectiveness of the SPT method is evaluated in a 50Gbaud dual-polarization 4SC-DSCM simulation system over 80km fiber. Fig. 5 gives the BER versus OSNR with different RSOP speeds. The BER curve of that without polarization crosstalk is also plotted as the benchmark. As shown in Fig. 5, there is almost no OSNR penalty compared to that without polarization crosstalk even though the RSOP speed is up to 10Mrad/s, which implies that the proposed SPT scheme effectively eliminates the polarization crosstalk even in an ultra-fast RSOP scenario. Figs. 6(a) to 6(d) give the spectrums of the received signals before and after polarization demultiplexing with RSOP of 10Mrad/s. It can be seen that the pilot tone inserted in the X-polarization will have a project in the Y-polarization due to the polarization rotation. After polarization demultiplexing based on the SPT scheme, the pilot tone in the Y-polarization disappears as shown in Fig. 6(d), indicating that the polarization crosstalk is eliminated. Then, the amplitude of the estimated $J_{XX}$ and $J_{YX}$ is displayed in Fig. 7, which is compared to the set one. The results suggest that the estimated polarization variation trend can nearly match the set one as it is the premise to achieve correct polarization demultiplexing. The aforementioned simulation results are obtained based on the SPT scheme without any frequency interval. To investigate whether reserving a certain frequency interval for the pilot tone will affect the transmission performance, the BER performance of the SPT scheme with different frequency intervals between two subcarriers is given in Fig. 8. According to the simulation results, the BER remains almost unchanged, even if the frequency interval is increased, and the BER is also similar to that without frequency interval given in Fig. 5. Therefore, the frequency interval can be removed to improve the spectral efficiency. Afterward, the BER performance of the proposed SPT scheme is also compared to the DPT and the conventional *N*-tap 2×2 CMMA methods under different RSOP conditions. The simulation results are shown in Fig. 9. It can be observed that the BER curves of the SPT and DPT schemes nearly overlap, exhibiting similar performance. Despite increasing the RSOP speed, the BER remains stable for these two schemes. However, the BER of the conventional *N*-tap 2×2 CMMA method becomes worse when the RSOP speed exceeds 100krad/s, and it fails to achieve polarization multiplexing when the RSOP speed reaches 1Mrad/s. Furthermore, the effect of the PDL on the proposed SPT scheme is also explored. Fig. 10 gives the BER curves versus PDL with different RSOP speeds, and the PDL is defined as equation (36). As shown in Fig. 10, the impact of the PDL can be neglected for the cases of RSOP speeds of 100krad/s and 1Mrad/s. It also introduces a slight impact on the performance of the signal with an RSOP speed of 10Mrad/s when the PDL reaches ±3dB. The results hint that the proposed SPT scheme is robust to the PDL.

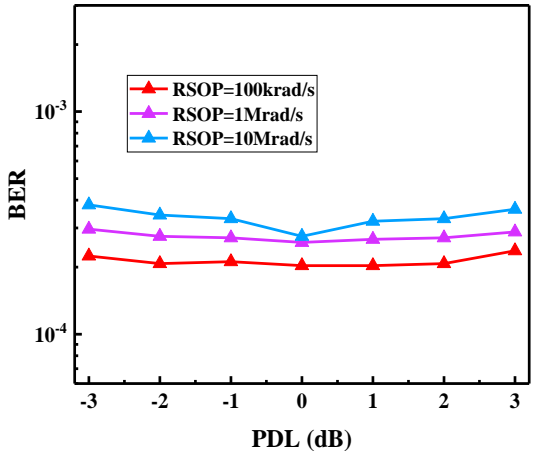

Fig. 10. BER curves versus PDL of the SPT method with different RSOP speeds.

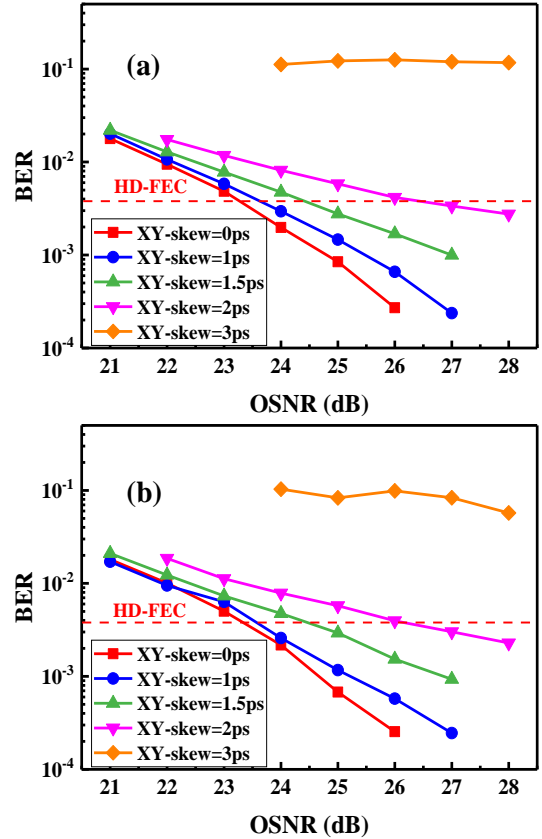

Fig. 11. BER versus OSNR of (a) the SPT method and (b) the DPT method under different Rx-XY-skew values.

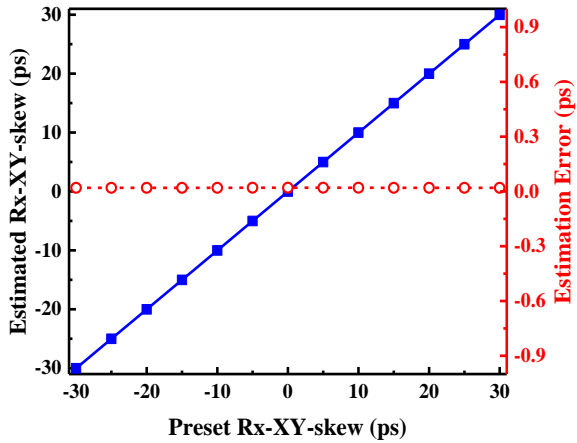

Fig. 12. Estimated Rx-XY-skew versus preset skew values of the proposed method.

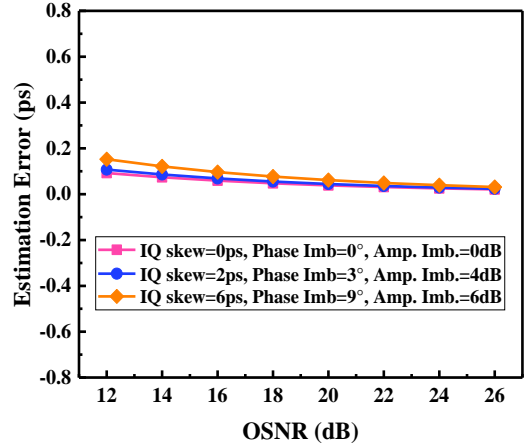

Fig. 14. BER curves of the SPT method with and without Rx-XY-skew compensation under different Rx-XY-skew values.

*C. Impact of the Rx-XY-skew*

According to the above simulation results, the proposed SPT method exhibits excellent performance in dealing with the polarization crosstalk under ultra-fast RSOP conditions, while it also improves the spectral spectrum. However, as discussed in section II-C, these PT-PDM schemes are susceptible to the Rx-XY-skew. In this part, the influence of the Rx-XY-skew in SPT and DPT methods is revealed by simulation. Figs. 11(a) and 11(b) show the BER performance of the SPT method and DPT method respectively, with different Rx-XY-skew values. As shown in Fig. 11(a), the larger the Rx-XY-skew, the worse the BER performance. Compared to the case without Rx-XY-skew, the OSNR penalties are 0.5dB, 1.2dB, and 3.2dB for the cases with Rx-XY-skew of 1ps, 1.5ps, and 2ps respectively at the hard-decision forward error correction (HD-FEC) threshold. As for the Rx-XY-skew reaching 3ps, the SPT method fails to eliminate the polarization crosstalk, even though increasing the system OSNR. The BER performance of the DPT method given in Fig. 11(b) is similar to that of the SPT method. It is also inoperative when the Rx-XY-skew reaches 3ps. The simulation results given in Figs. 11(a) and 11(b) reveal that the Rx-XY-skew indeed introduces significant performance degradation for these PT-PDM schemes, which is consistent with the analysis in section II-C. Therefore, the calibration of the Rx-XY-skew is essential for implementing the PT-PDM and the subsequent simplified equalization.

*D. Effectiveness of the proposed PT-MGPD scheme*

In this subsection, the effectiveness of the proposed PT-MGPD scheme for Rx-XY-skew estimation is evaluated by simulation. The estimation accuracy is investigated first. Fig. 12 shows the estimated Rx-XY-skew and the corresponding estimation error with different preset skew values. Note that, the frequency of the pilot tone is set to 2GHz in this simulation, and the simulation is conducted in the OBTB condition with OSNR of 26dB. The preset Rx-XY-skew values are swept from -30ps to 30ps. According to the simulation results, the estimation error of the proposed method can be less than ±0.1ps with a range from -30ps to 30ps. Then, the influence of the receiver-side IQ imbalance on the accuracy of the Rx-XY-skew estimation is also investigated. Figs. 13(a)-13(c) show the estimation error under the joint influence of Rx-skew and phase imbalance, Rx-skew and amplitude imbalance, phase imbalance and amplitude respectively. The preset Rx-XY-skew is fixed at 5ps, while the Rx-skew, phase imbalance, and amplitude imbalance are varied from -10ps to 10ps, -15° to 15°, and -10dB to 10dB. The simulation results indicate that the receiver-side IQ imbalance has almost no impact on the accuracy of the Rx-XY-skew estimation, since the estimation error is within 0.2ps with different degrees of receiver-side IQ imbalance. Furthermore, the robustness of the proposed method to the system OSNR is discussed. Fig. 14 gives the estimation error curves versus OSNR with and without the other receiver-side IQ imbalance. The results indicate the robustness of the proposed PT-MGPD method to OSNR since the estimation accuracy can be guaranteed even if the OSNR decreases to 12dB. Once the Rx-XY-skew is acquired, it can be calibrated before other DSP algorithms to ensure the effectiveness of the PT-PDM scheme as shown in Fig. 4. The BER performance of 50Gbaud 4SC-16QAM DSCM signal with and without Rx-XY-skew compensation is investigated for comparison. The simulation results are given in Fig. 15 with Rx-XY-skew varied from -3ps to 3ps, and the RSOP speed is 10Mrad/s. As shown in Fig. 15, for the case without Rx-XY-skew compensation, the BER performance deteriorates significantly as the Rx-XY-skew increases. However, the BER curve is flat with Rx-XY-skew compensation based on the estimated skew value and the performance is similar to that without Rx-XY-skew.

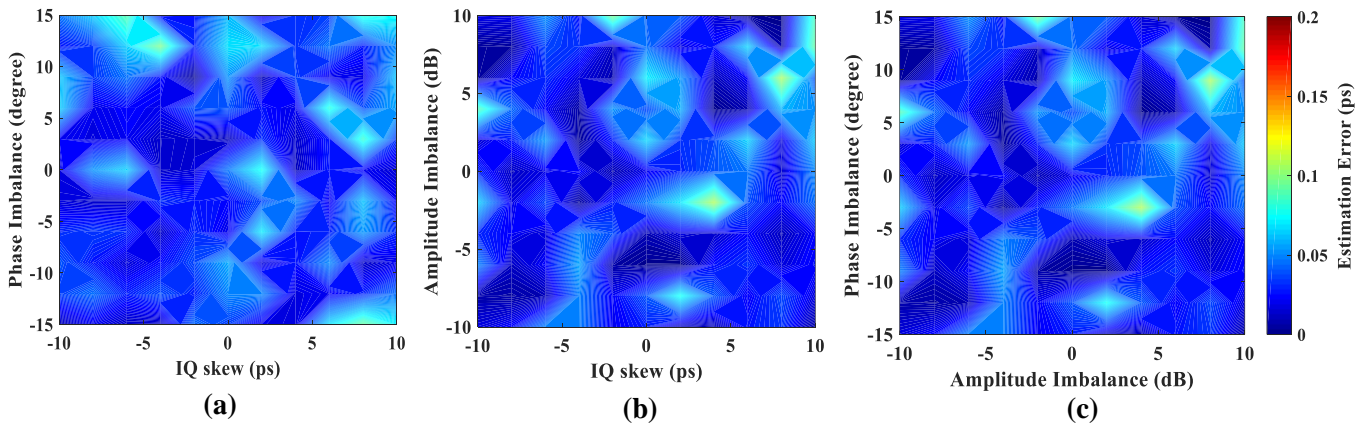

Fig. 13. Impact of receiver-side (a) IQ skew and phase imbalance, (b) IQ skew and amplitude imbalance, (c) amplitude imbalance and phase imbalance on the estimation accuracy of the proposed PT-MGPD scheme.

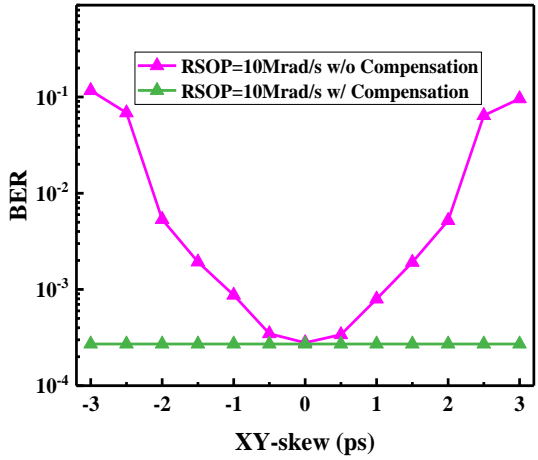

Fig. 15. BER curves of the SPT method with and without Rx-XY-skew compensation under different Rx-XY-skew values.

IV. EXPERIMENTAL SETUP AND RESULTS

*A. Experimental setup*

Then, the accuracy of the proposed PT-MGPD method for Rx-XY-skew estimation as well as the effectiveness of the SPT method are investigated experimentally. Fig. 16 gives the experimental setup. At the transmitter side, an external cavity laser (ECL) with linewidth less than 100kHz is used as the optical carrier, which is then injected into a DP-IQM. The power of the optical carrier emitting from the ECL is 16dBm. Before being launched into an 80km single-mode fiber (SMF), the modulated optical signal is amplified by an Erbium-doped fiber amplifier (EDFA). To adjust the OSNR of the system, a variable optical attenuator (VOA) and an EDFA are implemented to introduce different levels of amplified spontaneous emission (ASE) noise. At the receiver side, another ECL is served as the LO, which is injected into an integrated coherent receiver (ICR) together with the incoming signal for coherent detection. After that, a real-time oscilloscope (Tektronix MSO 72304DX) with a sampling rate and cut-off bandwidth of 50GSa/s and 20GHz is used to capture the detected signals. Finally, the captured signals are processed in the offline DSP. Both the Tx-DSP and Rx-DSP are achieved by MATLAB, and the corresponding DSP flows are the same as the simulation given in Fig. 4. Due to the limitations of the sampling rate and cut-off bandwidth of the oscilloscope, a dual-polarization 35Gbaud 4SC-16QAM signal is used for BER performance analysis in the experiment. As for the Rx-XY-skew estimation, the LO comes from the transmitter-side ECL to achieve the self-coherent detection, and a polarization controller (PC) is used to introduce proper polarization crosstalk.

*B. Transmission performance analysis*

The BER of the 35Gbaud 4SC-16QAM signal utilizing the SPT and DPT methods under different Rx-XY-skew conditions is measured first, and the experimental results are displayed in Figs. 17(a) and 17(b) respectively. Obviously, significant performance degradation can be observed for both the SPT and DPT methods as the Rx-XY-skew increases. As shown in Fig. 17(a), the OSNR penalties are about 1dB, 3dB, and 5.5dB at the HD-FEC threshold when experiencing 1ps, 2ps, and 3ps Rx-XY-skew compared to that without Rx-XY-skew impairment. As the Rx-XY-skew reaches 4ps, the BER can hardly be reduced despite increasing the OSNR, implying that the polarization crosstalk is unsolved in this situation. As for the results of the DPT method shown in Fig. 17(b), the situation is almost the same as the SPT method. The experimental results indicate that the Rx-XY-skew will seriously destroy the effectiveness of these PT-PDM schemes, which is not conducive to the simplification of the subsequent equalizer. Therefore, the compensation of the Rx-XY-skew is crucial.

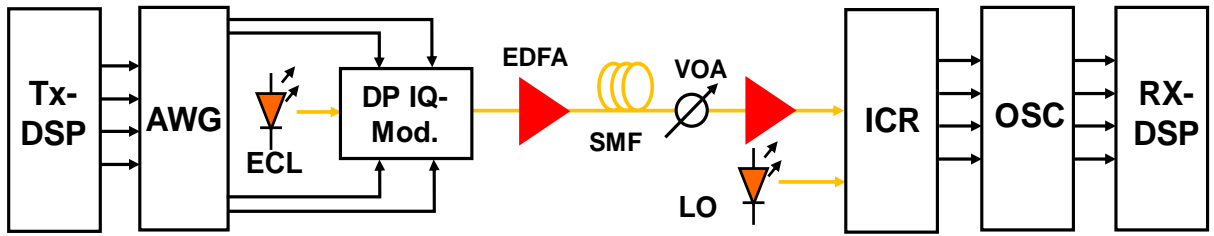

Fig. 16. Experimental setup of the transmission system.

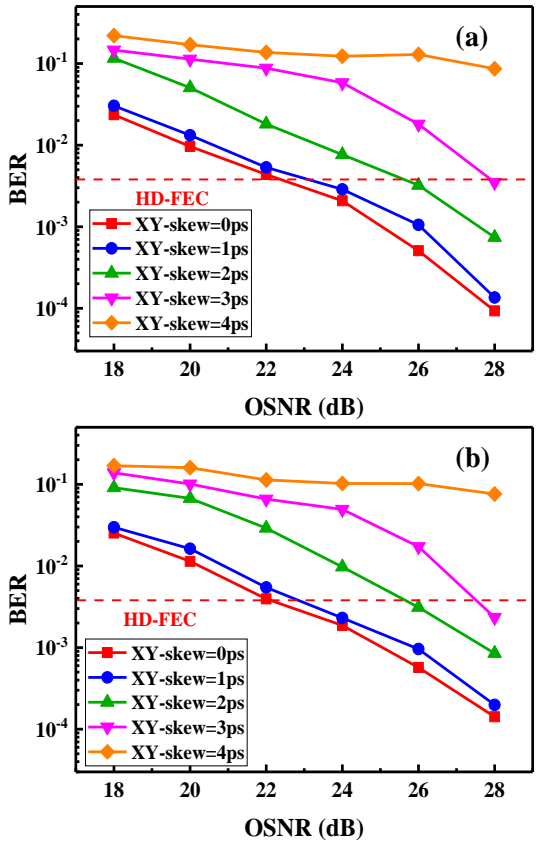

Fig. 17. Measured BER of (a) the SPT method and (b) the DPT method in the experiment under different OSNR conditions.

Then, the accuracy of the proposed PT-MGPD method based on a training signal composed of four frequency-and-polarization interleaved pilot tones is evaluated experimentally. Fig. 18(a) gives the estimated Rx-XY-skew under different preset skew values, and the corresponding estimation error is also plotted. According to the experimental results, the proposed PT-MGPD method can guarantee the estimation error within 0.3ps with range from -30ps to 30ps. Based on the estimated skew values, the Rx-XY-skew can be calibrated. Fig. 18(b) compares the BER performance of the 35Gbaud 4SC-16QAM signal utilizing the SPT method with and without Rx-XY-skew compensation. The BER is measured by sweeping the Rx-XY-skew, and the RSOP speed is set to 10Mrad/s. As for the case without Rx-XY-skew compensation, the BER increases rapidly as the Rx-XY-skew increases. In contrast, the BER curve is almost flat with Rx-XY-skew calibration based on the estimated skew value and achieves comparable performance as the case without Rx-XY-skew. Besides, insets (i) and (ii) give the constellations of each subcarrier carried on X-polarization and Y-polarization with 4ps Rx-XY-skew. It can be seen that the signals on different subcarriers are damaged to varying degrees, and some of them are even difficult to realize signal recovery. After the skew compensation, signals on all subcarriers for both polarizations have clear constellations as shown in insets (iii) and (iv). The experimental results indicate that the proposed MGPD method can effectively ensure the effectiveness of the PT-PDM scheme. After solving the Rx-XY-skew issue, the BER performance of these PT-PDM schemes is investigated and compared to the conventional $N$-tap 2×2 CMMA equalizer. The results are given in Fig. 18(c) with an OSNR of 28dB, which indicates that both the SPT and DPT methods offer significant advantages over the conventional CMMA equalizer in terms of tacking high-speed RSOP. The BER curves of these two PT-PDM schemes remain flat despite the RSOP speed reaching 10Mrad/s. In contrast, the BER of the signal utilizing the conventional CMMA equalization deteriorates rapidly when the RSOP speed exceeds 100krad/s. Therefore, the SPT and DPT methods not only have an enhanced capacity to address polarization crosstalk but also alleviate the computational complexity in comparison to the conventional CMMA equalizer, while the complexity can be further reduced by utilizing the SPT method as only a single pilot is used to achieve the polarization de-crosstalk.

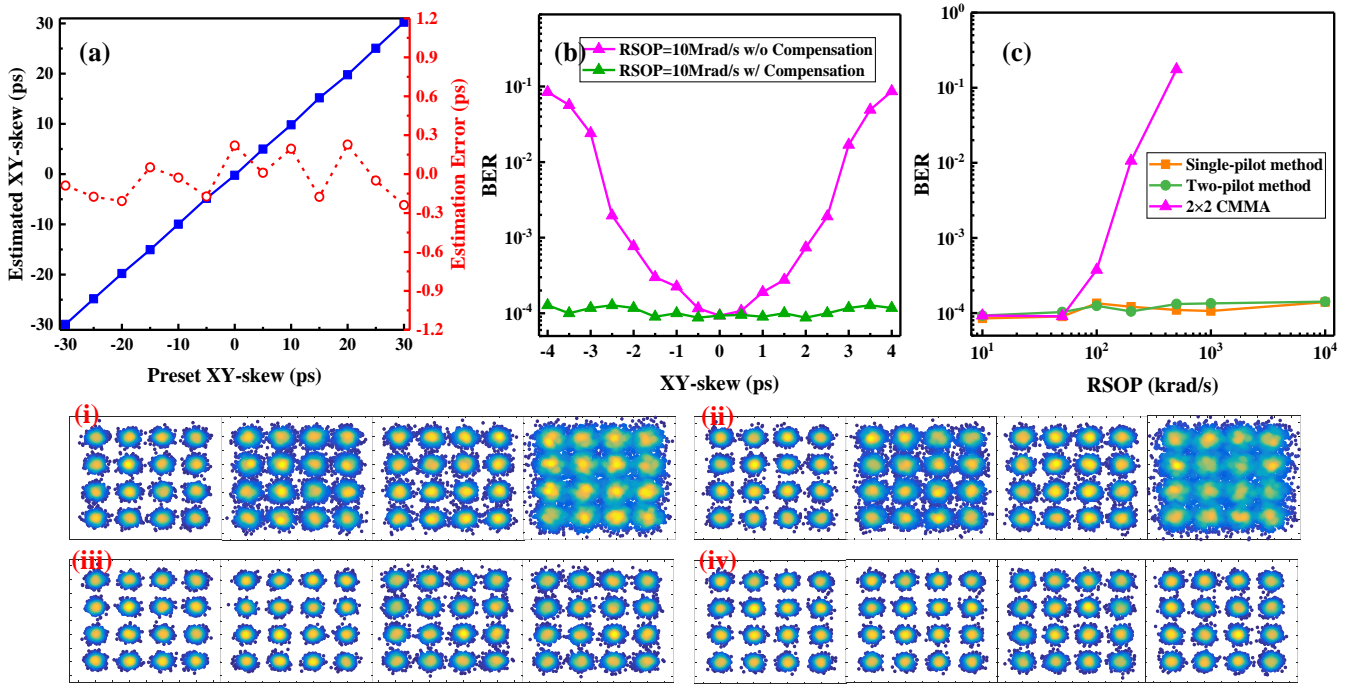

Fig. 18. (a) Estimated XY-skew and corresponding error versus preset XY-skew. (b) BER performance of the signal without and with Rx-XY-skew utilizing the SPT method. (c) Measured BER under different RSOP speeds in the experiment. Constellations of the signal with 4ps Rx-XY-skew: (i) X-polarization, (ii) Y-polarization. Constellations of the signal with Rx-XY-skew compensation: (iii) X-polarization, (iv) Y-polarization.

## V. Conclusion

PT-PDM has been proven to be a promising solution for polarization decoupling in coherent systems. It not only reduces the computation complexity of the coherent receiver effectively but also enhances the polarization rotation tracking speed. However, its effectiveness will be significantly compromised by the Rx-XY-skew impairment. In this paper, a PT-MGPD Rx-XY-skew estimation scheme based on a single-frequency-cosine pilot tone is proposed, serving as the prerequisite for the successful implementation of the PT-PDM and simplification of the adaptive equalizer. Both the simulation and experiment are conducted to prove the effectiveness of the proposed PT-MGPD method. According to the results, the proposed PT-MGPD method demonstrates reliable performance, achieving estimation errors within ±0.3ps. Furthermore, a modified low-complexity and high-spectral-efficiency polarization demultiplexing method named the SPT method is proposed in this work, which is only enabled by a single pilot tone and removes the guard band. The polarization demultiplexing performance of the proposed SPT method is investigated through simulation and experiment, which is also compared to the DPT method and the conventional $N$-tap 2×2 CMMA equalizer. According to the results, the proposed SPT method can achieve comparable performance to the DPT method. It successfully decouples the polarization crosstalk even with the RSOP speed reaching 10Mrad/s, which is impossible for the conventional CMMA equalizer. According to the results, the proposed schemes provide a hardware-efficient and reliable coherent DSCM solution for next-generation ultra-high-speed DCIs.

> REPLACE THIS LINE WITH YOUR PAPER IDENTIFICATION NUMBER (DOUBLE-CLICK HERE TO EDIT) <    12